# Audit Analysis Models, Security Frameworks and Their Relevance for VoIP

Oscar Gavilanez[1, 2], Glen Rodriguez[2] and Franklin Gavilanez[3]
[1]Systems Engineering, Escuela Superior Politécnica de Chimborazo ESPOCH, Ecuador
[2]Systems Engineering, Universidad Nacional Mayor de San Marcos UNMSM, Peru
[3]Mathematics, Montgomery College, USA
oscar.gavilanez@unmsm.edu.pe
glen.rodriguez@gmail.com
franklin.gavilanez@montgomerycollege.edu

**Abstract:** Voice over IP (VoIP) is the transmission of voice and multimedia content over Internet Protocol (IP) networks, this paper reviews models, frameworks and auditing standards proposed to this date to manage VoIP security through a literature review, with descriptions of both the historical and philosophical evolution reflecting an adequate knowledge of related research. Three research questions are raised here: RQ1. What are the requirements to be met by a model of security audit in VoIP systems to achieve their goals? RQ2. Today, are there additional attacks that previous works have not considered? RQ3. Which security requirements in the VoIP systems are covered (and which are not covered) by security frameworks? After some discussion about VoIP Protocols, Attacks on VoIP, Information Technology (IT) audit, IT security audits, Frameworks and auditing standards, we present a unified view of VoIP Security Requirements; as well as considering the contributions and disadvantages of frameworks and auditing standards toward achieving those requirements through a comparative evaluation. It was determined that there is no security framework which considers social engineering attacks in spite of being an important aspect to consider in security management VoIP; also there is no specific framework that covers all categories of security requirements for VoIP system, therefore, a more extensive model is needed.

**Keywords:** VoIP security, attacks on VoIP, IT security audit, IT security frameworks, social engineering attacks

## 1. Introduction

This article aims to analyse models, frameworks and auditing standards for relevance in VoIP systems; and to establish if previous jobs meet all safety requirements in VoIP systems. Three research questions are raised here: RQ1. What are the requirements to be met by a model of security audit in VoIP systems to achieve their goals? RQ2. Today, are there additional attacks that previous works have not considered? RQ3. Which security requirements in the VoIP systems are covered (and which are not covered) by security frameworks?

The paper is organized as follows. Section 2 shows the literature review regarding attacks on VoIP Security, IT security audits, framework and auditing standards, and VoIP Security Requirements. Section 3 deals with approaches and benchmarking of frameworks and standards, requirements and needs of the VoIP management systems. Finally, section 4 concludes with answers to the research questions.

## 2. Literature review

### 2.1 Introduction to VoIP

VoIP refers to communication protocols, technologies, methodologies and data transportation involved in the delivery of voice communications and multimedia sessions over the internet, rather than the public switched telephone network (PSTN). The communication voice uses the internet protocol (IP) which is low cost and is a major factor behind VoIP implementations, thus organizations can save thousands of dollars a year by opting for this technology, however, safety related problems should then be considered.

Ibrahim, Abdullah and Dehghantanha (2012) state that VoIP applications combine video and audio data along with the usual data packets traveling within a network system. Users make phone calls using softphones or IP phones (like Skype) and send instant messages through their computer. Geneiatakis, Lambrinoudakis and Kambourakis (2008) state that VoIP infrastructure inherits and uses various protocols of the stack architecture of the Internet, specifically in the network and transport use of Internet Protocol IP, and TCP, UDP or SCTP, respectively.

## 2.2 Attacks on VoIP security

An attempt to overcome the safety rules of a computer network is considered as an attack, based on a threat by exploiting vulnerabilities. Their effectiveness depends on the combination of three factors: countermeasures and security mechanisms used, existence of specific vulnerabilities on the infrastructure, and abilities of the attacker and tools used to perform the attack.

Shukla and Sahni (2013) addressed the detection and mitigation of SPIT using the signalling protocol analysis. Kshetri (2006) states that the constant reporting of vulnerabilities in information systems, and the exploit of human, procedural or technological failures on computing infrastructures in the world, are offering a perfect scenario for computer intrusions to grow.

Rosenberg et al (2002) argue that Session Initiation Protocol (SIP) has several security vulnerabilities, some are documented in RFC 3261 IETF, therefore there are several SIP security attacks.

*2.2.1 Username enumeration*

Dwivedi (2009) states that the attacker employs usernames consisting of valid account information registered in a VoIP network. This can be achieved by sniffing, when authentication is required between a user agent and a SIP server, the Uniform Resource Identifier (URI) standard SIP traversing the network in plain text, it can be sniffed backwards on the network, and as a switching network offers little protection, an attack can be made by Address Resolution Protocol (ARP) poisoning, and man in the middle to capture all URI within the local subnet.

*2.2.2 Man in the middle*

According to Xin (2007) when the attacker intercepts traffic SIP call signalling messages and traffic from the calling party to the called party, once the attacker has won the position calls can be hijacked. Man in the middle on the enemy gains the ability to read, insert and modify at will messages between two parties without either of them knowing that the link between them has been compromised, the attacker is able to observe and intercept messages between two victims.

*2.2.3 Eavesdropping*

Dwivedi (2009) considers eavesdropping the data transmitted between two or more points without being directly involved in it, thus signalling messages and audio streams intersect; usually it happens when implantation of VoIP is not planned properly and share the same physical medium to transmit voice and data.

*2.2.4 SIP password cracking*

According to Dwivedi (2009) SIP password cracking for a given user is usually done through brute force technique and the use of a dictionary, especially effective with weak passwords; there are methods and tools to break and crack the hashes in order to obtain a username and password to use the identity of the victim maliciously.

*2.2.5 Denial of service (DoS)*

Chen, E. (2006) considers DoS as one of the most alarming threats on the Internet. A DoS attack attempts to make a network node not available by flooding it with illegitimate packets, usurping their bandwidth and / or overloading node resources.

*2.2.6 VoIP spam and phishing*

Dwivedi (2009) considers that this type of attack involves a caller pretending to be a trustworthy organization and asks for confidential and critical information. This is an attack on the privacy of the data. The victim gives his/her personal data.

*2.2.7 Toll fraud*

Dwivedi (2009) refers to it as the ability to have unauthorized access to the VoIP service. It can be performed by manipulating the messages and configuration of VoIP signalling access components, allowing the hackers to sequester the system, entering directly on phones and copying the credentials and placing them on their own computer.

### 2.2.8 Spam over internet telephony (SPIT)

Rebahi and Sisalem (2005) state that the term spam is mostly used in the context of email, but it can also generally refer to any unsolicited communication. According to MacIntosh and Vinokurov (2005), in the context of VoIP networks, SPIT generally refers to massive amounts of unsolicited calls, and involves sending emails to users against their will. Each VoIP account has an IP address associated with it. Messages to thousands of IP addresses are sent which can carry viruses and spyware.

### 2.2.9 Social engineering

Snyder (2015) argues that social engineering stands in a unique place among hacking techniques due to its reliance upon deception rather than technology. These types of attacks are dangerous due to their high success rate and simplicity. According to Allsopp (2009), who is a professional penetration tester, social engineering refers to "obtaining of confidential or privileged information by manipulating legitimate sources" and that it "always requires some degree of deception". Nwogu and Odoh (2015) argue that vishing is a tool used by attackers to harvest customers' credentials. An attacker phones the victims and uses social engineering to trick them into divulging secret information such as credit card information.

Chiappetta et al (2013) report that VoIP threats have long been studied, their taxonomy divides threats in several macro-categories, which include the following: social threats, eavesdropping, intentional interruption and service abuse.

## 2.3 IT audit/ IT security audits

According to Weber (1998), an audit is the process of collecting and evaluating evidence to see whether the performance of the system to protect assets, maintains data integrity and operates effectively and efficiently in accordance with the objectives of the organization. Meanwhile, according to Arens, Elder and Beasley (2012) the audit is the accumulation and evaluation of evidence about information to determine and report on the degree of correspondence between the information and established criteria. Xia (2014) states that the information system audit records information system user activity in the behaviour of a mechanism; it is not only able to identify who has access to the system, but also how to use the recording system, so as to provide the basis for the post-process of security incidents.

Ariffin et al (2014) consider that Information audit is crucial in understanding the current state of an organization. By implementing an information audit, the needs, the available resources and the gaps are defined and reported to the top management of an organization, thus helping them to identify steps to be taken in order to improve the way they manage information to stay competitive in their business. Maciejewska (2014) considers that the dynamic development of the data communications industry in recent years increases the risk of auditing of financial statements, to some extent.

Kooper, Maes and Lindgreen (2011) argue that information technology has become a vital and integral part of many business activities and of the support, sustainability, and growth of enterprises. Business and IT departments must understand each other and make together the strategic/tactical plans needed for achieving goals of the organization.

IT security audits determine whether an information system and its management meet both the legal expectations of customer data protection and the company's financial standards against various security threats. Ryoo et al (2014) state that the organizations have used traditional IT audits to evaluate issues such as availability to authorized users and integrity and confidentiality in data storage and transmission.

A company needs to make explicit how specific security measures are linked to the relevant security objectives, and how they are meant to address possible risks concerning business processes and assets to be protected. Security engineering methodologies aim to enable the systematic design and implementation of security measures.

Burgemeestre, Hulstijn and Tan (2013) argue that security engineering requires a broader perspective. It integrates security requirements with functional requirements and constraints derived from other sources.

## 2.4 Frameworks and auditing standards

There are many IT-related management frameworks, standards and methodologies in use today. None of them, on their own, are standalone IT governance frameworks, but they all have a useful role in the efficient management of IT operations.

### 2.4.1 The CALDER-MOIR IT Governance framework

According to Calder (2016), CALDER-MOIR IT is designed to help to use all these overlapping and competing frameworks and standards, and also to deploy the best practice guidance contained in the international standard for IT governance, ISO/IEC 38500. The framework harmonizes the different frameworks and standards as a graphic classification linking the main topics such as business strategy, risk, IT strategy, operations and change management capabilities.

### 2.4.2 SAS 70 Type II

Statement on Auditing Standards No. 70 (SAS 70) is an internationally recognized standard audit focused on internal and external controls for a company that provides services. The report consists of a review by a certified independent auditing firm.

### 2.4.3 GLBA

Gramm-Leach-Bliley Act (GLBA) which is also known as the Law for the Modernization of the Financial Services 1999 provides limited protection against the sale of their private financial information privacy.

### 2.4.4 ISO 27000

Defines a framework for managing information security for organizations of any size. The ability to define a standard for information security guarantees proper implementation and mobility in different organizations, adapted to the specific needs of each, but all governed and controlled similarly. The ISO 27000 family of standards creates, manages and measures the Information Security Management System (ISMS) which is defined as a set of rules and policies for secure information management in an organization.

### 2.4.5 PCI DSS

Payment Card Industry Data Security Standard is developed by a committee formed by debit and credit card companies called Payment Card Industry Security Standards Council (PCI SSC) as a guideline to help organizations that process, store and / or transmit cardholder data, to ensure data, in order to prevent fraud involving payments.

### 2.4.6 EU privacy (or Directive 95/46 / EC)

This is a directive of the European Union adopted in 1995, which regulates the processing of personal data in the European Union. It is an important component of EU privacy and human rights standards. On January 25, 2012, the European Commission presented a draft (Regulation European Data Protection General) to replace the Data Protection Directive.

### 2.4.7 COBIT

Ridley, Young and Carroll (2004) state that Control Objectives for Information and Related Technology (COBIT) is an open standard that is being used increasingly by a wide range of organizations around the world. COBIT is the most suitable control framework to help an organization to ensure alignment between the use of Information Technology (IT) and business goals, and emphasizing the need of business so that managers are satisfied by each objective of control. While there is a wide range of frameworks, standards and related IT control documents, the main focus of COBIT is in alignment with the use of IT in achieving organizational goals.

Relatively little academic literature published investigates the use of COBIT. This may be due to extensive electronic resources available on COBIT being primarily designed for IT and audit professionals. These sources are produced by the Information Systems Audit and Control Association (ISACA) and the IT Governance Institute and are not known by many academic authors. There are few published studies concerning the adoption or use of COBIT. Apart from the study cases produced by the IT Governance Institute, there is little literature that considers the scope and characteristics of organizations that have used COBIT and the results of the application.

*2.4.8 The Information Technology Infrastructure Library (ITIL) framework*

ITIL is a set of practices for IT service management that focuses on aligning IT services with the needs of business; despite stating the risk should be identified, measured and mitigated, ITIL does not clarify what concrete method is to be used in dealing with the risk. The risk management component is implicit in the process but in a very abstract, unobvious way and it is not specified as desired in the actual business scenario.

Vilarinho and da Silva (2011) consider that the risk management is not clearly shown because there is not an obvious way to implement risk management in ITIL, despite risk management being referenced in some of the ITIL books, mainly in Operation and Continual Service improvement.

Goldstein and Frank (2016) state that there is a need for a method to guide the design of IT security systems, balancing the goals and limitations of various perspectives. Burgemeestre, Hulstijn and Tan (2010) manifest that in business administration, high-level control models such as Committee of Sponsoring Organizations (COSO), Simons's levers of control, and IT governance models such as COBIT, serve as inspiration for designing internal control systems. With regards to the design of computerized workplaces Draxler and Stevens (2011) found such permeation in the common standards for IT management like ITIL and COBIT, the ITIL standard does not address the single user as the customer of an IT service, but the organization as a whole.

## 2.5 VoIP security requirements

Bellavista, Corradi, and Stefanelli (1999) state that without security, systems cannot deal with global untrusted environments, such as the internet; without interoperability, they cannot interact with existing tools and legacy systems. The increasing complexity of global distributed systems, from a set of resources connected by network infrastructures to a set of network-centric coordinated services, has motivated the evolution of traditional management models. Dabbebi, Badonnel and Festor (2015) consider that telephony over IP has been widely deployed, supported by the standardization of VoIP signalling and media transfer protocols. This deployment has also led to the emergence of several security threats, including attacks inherited from the IP layer and attacks specific to the application layer. A large variety of security mechanisms has been proposed for addressing them, but these mechanisms may seriously degrade such a critical service. Telephony over IP defines a new paradigm, which enables the establishment and the transmission of voice communications directly over IP networks. This integration of data and voice on the same network enables higher flexibility and cost sharing. It also poses security challenges, as the VoIP service is much less confined than traditional PSTN. The risk is the combination of the probability that a given threat will exercise a vulnerability on a given system and the resulting impact of that adverse event on this system.

Zhang and Fischer-Hübner (2013) state that the transmissions of VoIP flows are sensitive to the quality of service (QoS). Three issues are often taken as evaluation criteria: end-to-end delay, delay jitter and packet loss. Several basic security requirements for VoIP systems are consolidated in Table 1.

**Table 1:** Basic security requirements for VoIP systems

| Requirements | Description |
|---|---|
| Authentication, integrity, confidentiality, privacy, efficiency, performance, security against various attacks, security features | Security measures are usually inversely proportional to the performance. The original SIP authentication scheme does not provide mutual authentication and cannot support the integrity and confidentiality protection. Special safety requirements for privacy protection, which have not been considered in most previous works include strong authentication, protection of privacy and efficiency. The authentication scheme for VoIP must meet several requirements of safety and efficiency, in order to meet several objectives, including: providing security against various attacks, providing security elements and generate privacy protection. |
| Quality of Service QoS | The transmission of VoIP flows is sensitive to the quality of service (QoS), three issues are often taken as evaluation criteria: end-to-end delay, delay jitter and packet loss |

Gonzalez et al (2012) carry out a classification of several potential security issues determining the following categories: network security, interfaces, data security, virtualization, governance, compliance and legal issues.

*Oscar Gavilanez, Glen Rodriguez and Franklin Gavilanez*### 3. Approaches and benchmarking

There are disadvantages to be taken into account with respect to approaches found on existing frameworks according to the literature; they justify the need for a model of Security Audit for VoIP; this are shown in Table 2.

**Table 2:** Disadvantages of frameworks and standards

| Author | Disadvantages of frameworks and standards |
|---|---|
| Cots, Casadesús and Marimon (2016) | The establishment of audit is a specific benefit of management in accordance with rules and cannot be extended to other frameworks like ITIL |
| Vilarinho and da Silva (2011) | ITIL has some gaps in the specification Risk Management |
| Goldstein and Frank (2016) | Management frameworks, such as ITIL and COBIT are not sufficient for the design and management of comprehensive IT security systems because they do not address the challenges in terms of protection of IT resources |
| Burgemeestre et al (2010) | The methods of risk management support decisions on the implementation of control measures, however, they are not designed as a means to explain and encourage compliance with decisions to external auditors. |
| Draxler and Stevens (2011) | The ITIL standard does not address the needs of a single user of an IT service but that of the organization as a whole, therefore, the issue of adaptation is not addressed in the ITIL standard (by the end users themselves or in cooperation with the service provider) |
| Burgemeestre, Hulstijn and Tan (2013) | The safety standards of information as COBIT or NIST 800-53 are organized around a set of objectives of generic control; open standards must be translated into specific rules to be applied. |
| Ridley, Young and Carroll (2004) | COBIT is an open standard that ensures alignment between the use of IT and business objectives; it stresses the need of business by each control objective; the IT control frameworks are designed to promote effective IT governance. |

**Table 3**: Contains requirements and needs of management systems VoIP versus some frameworks

| Framework | Management systems VoIP |
|---|---|
| COBIT sets best practices for IT controls, but companies should determine who controls make for your specific organization. That makes a very general framework COBIT The user must implement COBIT: Ensure that the application meets its regulatory requirements. Selecting the right controls for your organization. Monitoring and reporting on the program. | Integrated security management is required. Should identify, implement and automate the critical controls. Define, manage and report on a coherent set of internal controls over data and corporate systems. |
| The United States Government Configuration Baseline (USGCB) (formerly FDCC) and Security Content Automation Protocol (SCAP). Standard designed to provide a unique setting, consistent throughout the company, both for desktop and notebook PCs, and therefore reduce the costs associated with support, application compatibility and at the same time to improve safety. SCAP is based on a set of open standards that list the security issues configuration, software failures and product names. It is used for measuring systems to determine the presence of known vulnerabilities and provide a mechanism for sorting the results in order to evaluate the potential impact. | Automated systems evaluation is required on VoIP. Build the capacity to safely manage the configuration of Group Policy objects for VoIP. Integrated security management for compliance and policies to ensure IT assets and manage risk. Effective configuration management of all VoIP devices and identifying those where the system differs from its expected configurations. |
| Federal Information Security Management Act (FISMA) / NIST 800-53. Streamlining business processes to ensure business continuity, improve operational efficiency and maximize the security of IT infrastructures of organizations. They must implement strategies and processes to: assure service levels, policy compliance and appropriate risk management, reduce the cost and complexity of managing heterogeneous IT infrastructures. | Monitor heterogeneous security controls across the organization, enabling rapid identification of potential and existing threats and provide a detailed and accurate safety personal knowledge to allow quick recovery and reduce exposure times. Policies of real-time monitoring and change auditing (reports). Generate accurate and timely assessments of security risks. Management of user identity access and rights. |

148

| Framework | Management systems VoIP |
|---|---|
| Gramm–Leach–Bliley Act (GLBA). Financial services providers must protect customer information against threats to security, confidentiality and integrity; however, GLBA is not a technical safety standard. | Establishment of a program of information security to safeguard user information. The evaluation of the ability of security policies to identify and control internal and external threats. The management and control of risks that threaten customer information Establish Security Policy. Policies needed to address access controls, the use of encryption, exchange controls, monitoring and incident response. Risk Assessment Information Security and Configuration Management. Access controls systems. Real-time monitoring and auditing Reports of security breaches, audit for security applications and devices, as well as network devices. |
| ISO 27002 is an international safety standard or code of good practices for managing information security. The key to success lies in identifying a common framework for implementation and mapping regulatory requirements that framework, because the objective of ISO 27002 is to provide a comprehensive security framework, their requirements are very broad in its impact usually affecting all aspects of an IT organization. | Develop a set of policies, standards and other internal security guidelines. Access control. Management of rights of user identity and access, streamline password management, access control to sensitive information inside or outside the firewall. Management service levels. Protection against intrusions, security event management. Reporting on safety deficiencies. |
| In order to ensure the reliability of power generation and power transmission, North American Electric Reliability Corporation (NERC) Critical Infrastructure Protection (CIP) requires a wide range of both technical and procedural controls to be implemented and documented. NERC CIP involves specific technical security controls and process controls around training, policy enforcement, and access. | Secure record management, event correlation and threat detection and Safety Information. Evaluate default security. User Management. Integration of security and compliance tools. |

## 4. Conclusions

- Responding to the research question 1, we can say that the requirements to be met by a model of VoIP audit to achieve its objectives should be based on the following categories: network security, interfaces, data security, virtualization, governance, compliance and legal matters, adaptability to consolidated VoIP systems and social engineering.

- Regarding the research question 2, we consider that unlike common technical attacks, social engineering attacks cannot be prevented by current security tools and software. Despite the devastating nature of social engineering attacks, there seems to be a lack of concern about social engineering in VoIP professional literature, with writers and researchers devoting their time solely to technical security issues. Despite this lack of concern, social engineering remains perhaps the most dangerous threat to information security for any company; accordingly, social engineering continues to be an issue that must be addressed.

- Responding to the research question 3, according to the literature review there is no specific framework that covers all categories of security requirements for VoIP systems; a more extensive model is needed.

- Social engineering refers to taking advantage of human element of security to compromise vital information, when the ease of use and cost of execution of social engineering are compared to more technical attacks on companies, it becomes apparent why social engineering is such a common tool, it must pay careful attention to both technical security breaches and non-technical forms of hacking like social engineering.

- Security is a people and management problem; therefore, physical and technical controls are not enough to protect the security of the information.

*Oscar Gavilanez, Glen Rodriguez and Franklin Gavilanez*